\documentclass{PoS}

\rightline{\sffamily RBRC-1225}\vspace*{-10mm}

\title{Matching issue in quasi parton distribution approach}

\ShortTitle{Matching issue in quasi parton distribution approach}

\author{\speaker{Tomomi Ishikawa}\\
        T. D. Lee Institute,
	Shanghai 200240, China\\
	RIKEN BNL Research Center,
        Brookhaven National Laboratory, Upton, New York 11973, USA\\
        E-mail: \email{tomomi.ik@gmail.com}}

\author{Yang-Qing Ma\\
        School of Physics and State Key Laboratory of Nuclear Physics
	and Technology, Peking University,
	Beijing 100871, China\\
	Center for High Energy physics, Peking University,
	Beijing 100871, China\\
	Collaborative Innovation Center of Quantum Matter,
	Beijing 100871, China\\
        E-mail: \email{yqma@pku.edu.cn}}

\author{Jian-Wei Qiu\\
        Theory Center, Thomas Jefferson National Accelerator Facility,
        Newport News, VA 23606, USA\\
        E-mail: \email{jqiu@jlab.org}}

\author{Shinsuke Yoshida\\
	Theoretical Division, Los Alamos National Laboratory,
	Los Alamos, NM 87545, USA\\
	Physics Department, Brookhaven National Laboratory,
        Upton, New York 11973, USA\\
	Theoretical Research Division, Nishina Center, RIKEN,
        Wako, Ibaraki, 351-0198, Japan\\
        E-mail: \email{shinyoshida85@gmail.com}}	

\abstract{
In recent years, the quasi parton distribution has been introduced
for extracting the parton distribution functions from lattice QCD simulations.
The quasi and standard distribution share the same perturbative collinear 
singularity and the renormalized quasi distribution can be factorized into 
the standard distribution with a perturbative matching factor.
The quasi parton distribution is known to have power-law UV divergences,
which do not exist in the standard distribution.
We discuss in this talk the nonperturbative renormalization scheme for
the power divergence.
We also demonstrate the perturbative matching of the quasi quark distribution 
between continuum and lattice at the one-loop.
}

\FullConference{34th annual International Symposium on Lattice Field Theory\\
		24-30 July 2016\\
		University of Southampton, UK}

\begin{document}

\section{Introduction}

Quantum chromodynamics (QCD) covers a wide range of scales from
its partonic degrees of freedom, quarks and gluons, to complex
hadrons, such as pion and nucleon.
QCD is an asymptotic free theory, and a perturbative method could be 
applicable for studying observables with a large momentum transfer 
in high energy collisions.
On the other hand, at a low energy, the strong interaction physics is
nonperturbative.
When we study high energy scattering processes with identified
hadron(s), such as
deep inelastic scattering and Drell-Yan process,
one of the key concepts is the ``QCD collinear factorization''.
The scattering cross sections can be approximately written in a
convolution of perturbative hard part and nonperturbative parton
distribution functions (PDFs), which absorb all parturbative collinear
divergences of the partonic scattering.
The PDFs are universal functions and can be used to predict 
the cross sections of various hadronic scattering processes.

The PDFs have been extracted through global QCD analyses.
Direct calculation of the PDFs is, in principle, possible
and would give us invaluable insights into QCD dynamics,  
complementary to the global QCD analysis.
Lattice QCD is a possible nonperturbative method to calculate the PDFs.
However, since the PDFs are defined by using field operators located on
the light-cone, e.g.,
\begin{eqnarray}
q(x,\mu)=\int\frac{d\xi^-}{4\pi}e^{-ix\xi^-P^+}
\langle P|\overline{\psi}(\xi^-)\gamma^+
\exp\left(-ig\int_0^{\xi^-}d\eta^-A^+(\eta^-)\right)\psi(0)|P\rangle,
\label{EQ:normal-quark-PDFs}
\end{eqnarray}
for a quark distribution with the nucleon momentum $P=(P_0, 0, 0, P_z)$,
where $x$ is the momentum fraction of $P$ carried by the quark, 
$\mu$ is the factorization scale, and the light-cone coordinate 
$\xi^{\pm}=(t\pm z)/\sqrt{2}$,
the time-dependence of the fields correlated in the $\xi^-$-direction
makes the direct calculation on the Euclidean lattice impossible.
Although there have been attempts to calculate the moments of PDFs
on the lattice, and then reconstruct PDFs from the moments, 
this approach has not been very successful since the higher moments 
are noisy and the existence of power divergence causes complicated
operator mixings.

A recent breakthrough in the lattice calculation of the PDFs is
the quasi-PDF approach, introduced by Ji~\cite{Ji:2013dva}.
The quasi-PDFs are defined with fields correlated completely 
along the spatial direction, e.g.,
\begin{eqnarray}
\widetilde{q}(\tilde{x},\mu, P_z)=\int\frac{d\delta z}{4\pi}
e^{-i\delta z\tilde{x}P_z}
\langle P_z|\overline{\psi}(\delta z)\gamma^3
\exp\left(-ig\int_0^{\delta z}dz'A_3(z')\right)\psi(0)| P_z\rangle,
\label{EQ:quasi-quark-PDFs}
\end{eqnarray}
for the quasi-quark distribution, and are calculable on the Euclidean lattice.
The quasi-PDFs could be matched to normal PDFs using the large momentum
effective theory~\cite{Ji:2014gla} with perturbative matching factors
\begin{eqnarray}
 \widetilde{q}(x,\tilde{\mu}, P_z)=
  Z\left(x,\frac{\tilde{\mu}}{P_z}, \frac{\mu}{P_z}\right)\otimes q(x,\mu)
  +{\cal O}\left(\frac{\Lambda_{\rm QCD}^2}{P_z^2},
            \frac{M^2}{P_z^2}\right),
\end{eqnarray}
where $\otimes$ represents a convolution with respect to $x$
and $M$ is a nucleon mass.  
The relation between the normal- and renormalized 
quasi-PDFs was also investigated in terms of the QCD collinear
factorization approach \cite{Ma:2014jla, Ma:2014jga}.

While there have been several lattice calculations of quasi-PDFs 
using the matching approach introduced by Ji
~\cite{Chen:2016utp, Alexandrou:2016jqi}, a couple of uncertainties
remains unsolved in their simulations, e.g., the existence of power
divergences and matching between continuum and lattice.

In this article, we report our approach toward resolving these uncertainties.
We propose a non-perturbative renormalization of the quasi-PDFs to 
subtract their power divergences.
With our renormalization scheme, we provide an example of one-loop
perturbative calculation of the matching factor between continuum and lattice.

\section{Renormalization of a non-local operator with power divergence}

The quasi-quark distribution (\ref{EQ:quasi-quark-PDFs}) is known to have
the linear power divergence, which only comes from the Wilson line in its
definition.
If we adopt a UV cutoff as a regulator, the power divergence is manifest.
Since the lattice QCD naturally introduces the UV cutoff, the UV
divergence must be handled. Otherwise, we cannot take the continuum limit.

The renormalization of a Wilson line along a (smooth) contour $C$,
$W_C$, has been known to be
\begin{eqnarray}
 W_C=Z_ze^{\delta m \ell(C)}W_C^{\rm ren},
\label{EQ:wilson-line_renormalization}
\end{eqnarray}
where a superscript ``ren'' indicates the operator is renormalized,
$\ell(C)$ is length along the contour $C$,
and $\delta m$ depicts mass renormalization of a test particle moving
along the contour $C$
~\cite{Dotsenko:1979wb, Arefeva:1980zd, Craigie:1980qs,Dorn:1986dt}.
The power divergence is contained in the $\delta m$ in the exponential factor,
leaving only logarithmic divergences in the renormalization constant, $Z_z$, 
which arise from end points of the Wilson line. 
For the non-local quark bilinears of the hadronic matrix element 
in the r.h.s. of equation (\ref{EQ:quasi-quark-PDFs}), named as $O_C(z)$,
it was assumed to be~\cite{Dorn:1986dt}
\begin{eqnarray}
O_C=Z_{\psi,z}e^{\delta m \ell(C)}O_C^{\rm ren},
\label{EQ:non-local-bilinear_renormalization}
\end{eqnarray}
where $Z_{\psi, z}$ does not contain the power divergence
and $\delta m$ in the exponential factor,
which, like that in equation (\ref{EQ:wilson-line_renormalization}),
takes care of all the power divergence.
Unlike the Wilson line case in (\ref{EQ:wilson-line_renormalization}), 
the multiplicative renormalization pattern
(\ref{EQ:non-local-bilinear_renormalization}) is non-trivial.
While the renormalization pattern on the power divergence
in equation (\ref{EQ:non-local-bilinear_renormalization}),
which is in an exponential form, holds even nonperturbatively,
there is no guarantee on whether other divergences can be
multiplicatively renormalized~\cite{IMQY2}.

If we rewrite the Wilson line operator as an auxiliary fermion field
propagator, which is similar to a static quark propagator with the
field propagating in the $z$-direction, we can use the knowledge 
of the heavy quark effective theory (HQET).
In the HQET case, the multiplicative renormalizability has been
seen up to first several loops.
Lattice QCD simulation on the HQET also suggests the nonperturbative
renormalizability, because the existence of the continuum limit of
the heavy-light system has been numerically checked.
Therefore we have a good reason to assume the renormalization pattern
(\ref{EQ:non-local-bilinear_renormalization}) and use it
in following arguments.
Knowing the power divergence can be renormalized in the exponential form
as in equation (\ref{EQ:non-local-bilinear_renormalization}),
the power divergence in the quasi-quark distribution could be subtracted
by introducing a non-local operator in the hadronic matrix element
of the quasi-quark distribution in (\ref{EQ:quasi-quark-PDFs})
~\cite{Ishikawa:2016znu, Chen:2016fxx}:
\begin{eqnarray}
O^{\rm subt}(\delta z)=e^{-\delta m|\delta z|}\overline{\psi}(\delta z)\gamma^3
P\exp\left(-ig\int_0^{\delta z}dz'A_3(z')\right)\psi(0),
\label{EQ:subtracted_non-local_operator} 
\end{eqnarray}
where the superscript ``subt'' indicates that this is a power divergence
subtracted operator.

We now need some scheme to fix the mass renormalization $\delta m$.
One of the convenient choices for the subtraction scheme is
to use a static quark potential $V(R)$,
which shares the same power divergence as the one in the non-local operator
~\cite{Musch:2010ka}.
The static potential $V(R)$ can be obtained from an $R\times T$ Wilson loop
in the large $T$ limit:
\begin{eqnarray}
W_{R\times T}\propto e^{-V(R)T}\;\;\; (T\rightarrow {\rm large}). 
\end{eqnarray}
The renormalization of the static potential is written as
\begin{eqnarray}
V^{\rm ren}(R)=V(R)+2\delta m. 
\end{eqnarray}
To fix $\delta m$, we impose a fixing condition at some distance $R_0$,
yielding
\begin{eqnarray}
V^{\rm ren}(R_0)=V_0\longrightarrow
\delta m=\frac{1}{2}\left(V_0-V(R)\right).
\label{EQ:subtraction_condition}
\end{eqnarray}
Since the Wilson loop is measured in the lattice simulation,
the subtraction of the power divergence in equation
(\ref{EQ:subtracted_non-local_operator}) is nonperturbative.

\section{One-loop perturbation contribution in the continuum}

\begin{figure}[t]
\vspace*{-5mm}  
\begin{center}
\parbox{34mm}{
\begin{center}
\includegraphics[scale=0.28, viewport = 0 0 320 300, clip]
{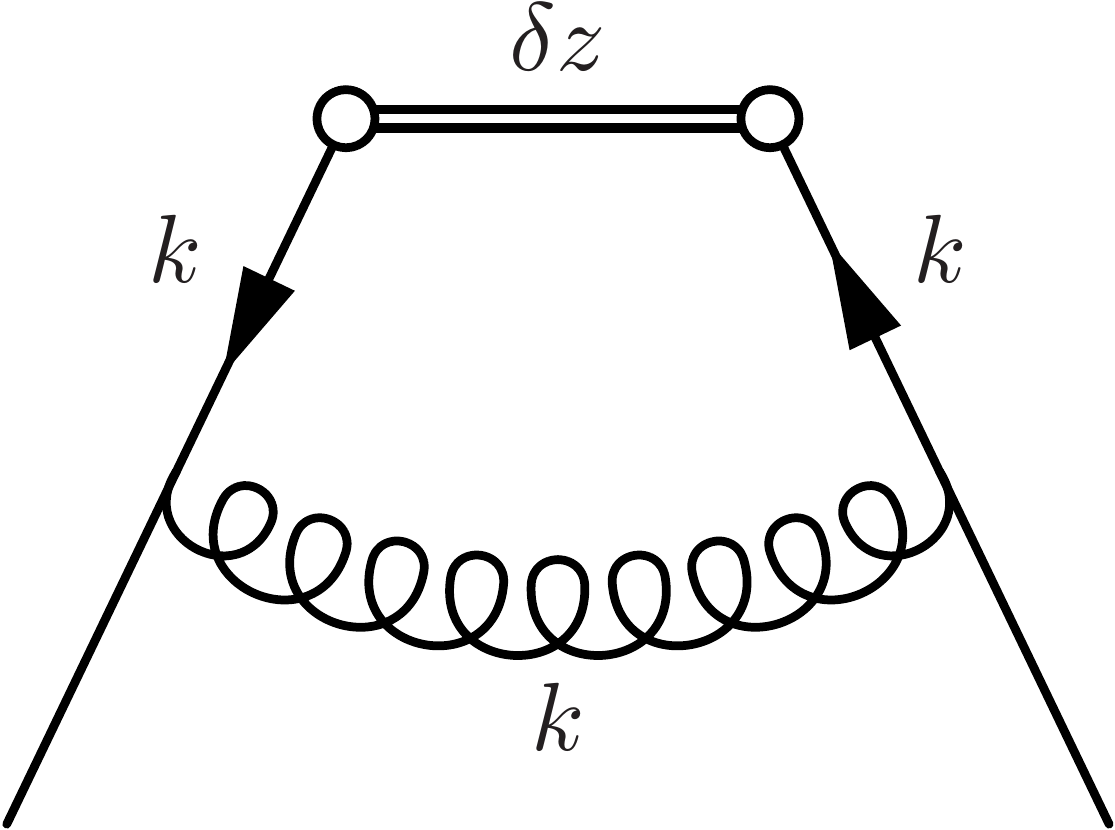}
$\delta\Gamma_{\rm vertex}(\delta z)$
\end{center}
}
\parbox{76mm}{
\begin{center}
\includegraphics[scale=0.28, viewport = 0 0 320 300, clip]
{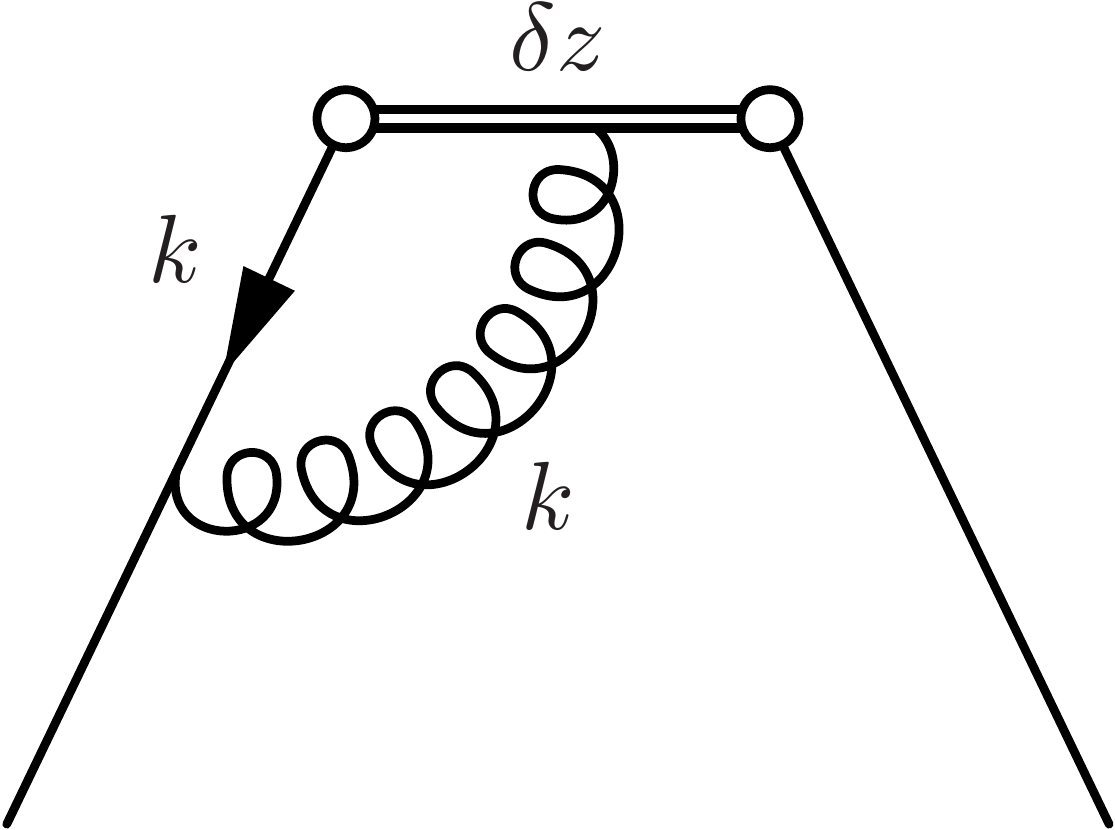}
\includegraphics[scale=0.28, viewport = 0 0 320 300, clip]
{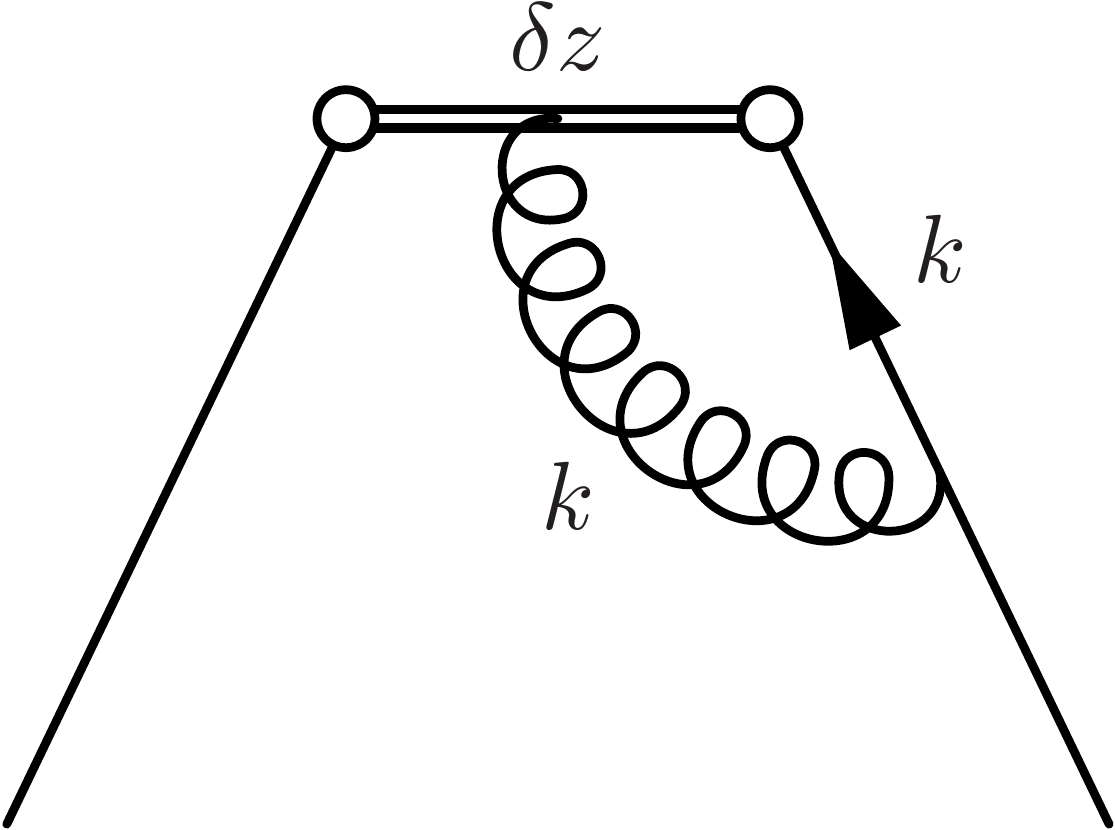} 
$\delta\Gamma_{\rm sail}(\delta z)$
\end{center}
}
\parbox{34mm}{
\begin{center} 
\includegraphics[scale=0.28, viewport = 0 0 320 320, clip]
{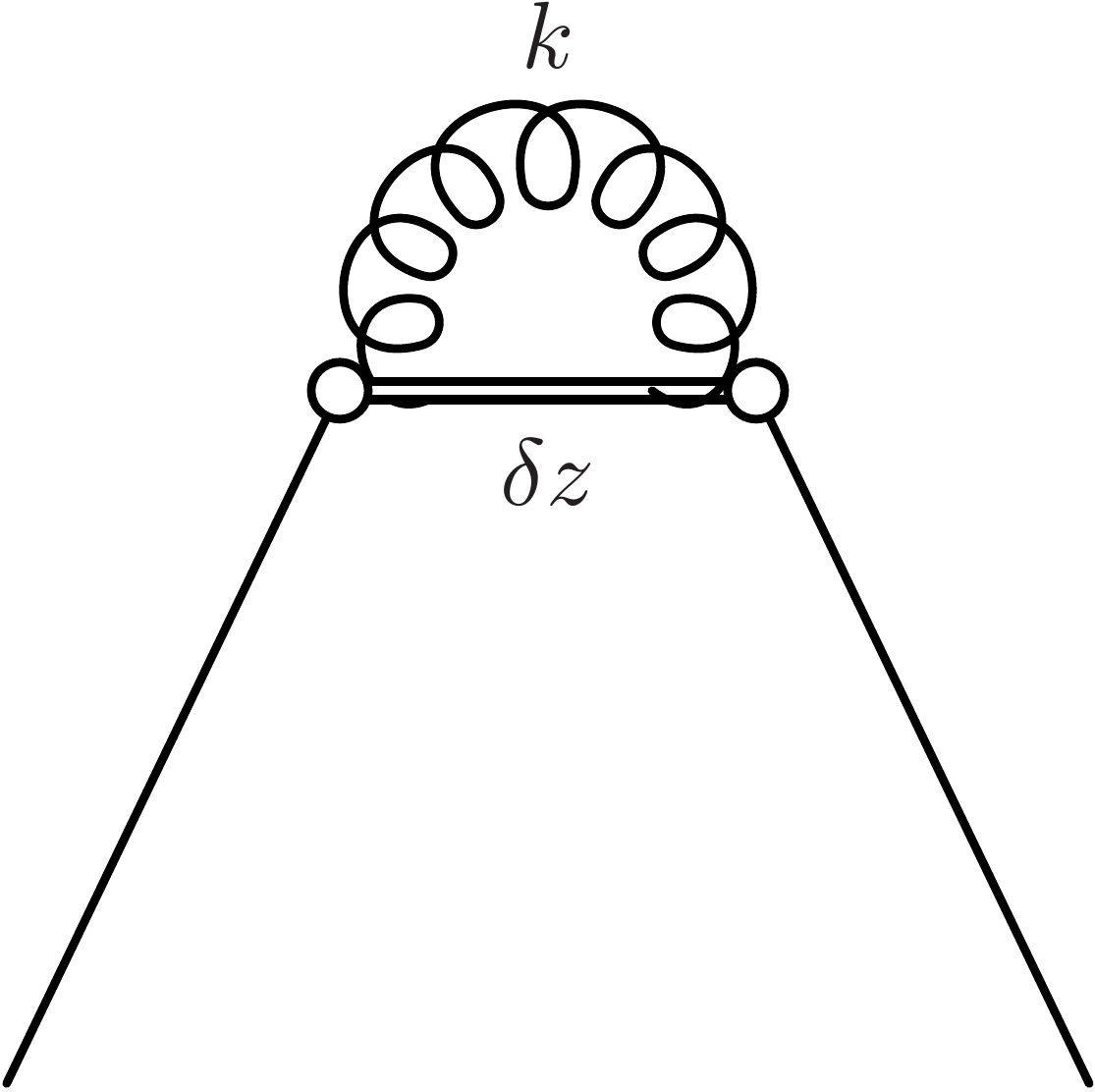} 
$\delta\Gamma_{\rm tadpole}(\delta z)$
\end{center}
 }
 \vspace*{-2mm}
\caption{One-loop diagrams.}
 \label{FIG:feynman_diagram_deltaGamma} 
 \end{center}
\vspace*{-5mm}
\end{figure}

In this section, we present an one-loop calculation of the matrix
element in the r.h.s. of equation (\ref{EQ:quasi-quark-PDFs}) in the continuum.
In this calculation, we set external quark momenta to be zero,
because the results are used for obtaining the matching factor between
the continuum and the lattice calculation,
and the external momentum dependence is canceled in the matching.
We assume Euclidean space in the calculation.
Besides the wave function renormalization of quarks, there are three
diagrams to be calculated at the one-loop level in the Feynman gauge, 
as shown in figure \ref{FIG:feynman_diagram_deltaGamma}.
By integrating out $z$ component of the loop momentum, we obtain
\begin{eqnarray}
\delta\Gamma_{\rm vertex/sail/tadpole}(\delta z)
=\frac{g^2}{(4\pi)^2}G_F\gamma_3I_{\rm vertex/sail/tadpole}(\delta z),
\end{eqnarray}
\begin{eqnarray}
I_{\rm vertex}(\delta z)&=&
\frac{(4\pi)^2}{4}\int_{k_{\perp z}}
\left(\frac{1}{k_{\perp z}^3}+\frac{|\delta z|}{k_{\perp z}^2}
+\frac{|\delta z|^2}{k_{\perp z}}\right)e^{-k_{\perp z}|\delta z|},
\\
I_{\rm sail}(\delta z)&=&
\frac{(4\pi)^2}{2}\int_{k_{\perp z}}
\left[\frac{1}{k_{\perp z}^3}
-\left(\frac{1}{k_{\perp z}^3}+\frac{|\delta z|}{k_{\perp z}^2}\right)
e^{-k_{\perp z}|\delta z|}\right],
\\
I_{\rm tadpole}(\delta z)&=&
\frac{(4\pi)^2}{2}\int_{k_{\perp z}}
\left[\frac{1}{k_{\perp z}^3}-\frac{|\delta z|}{k_{\perp z}^2}
-\frac{1}{k_{\perp z}^3}e^{-k_{\perp z}|\delta z|}\right],
\end{eqnarray}
where $C_F=4/3$ and $k_{\perp z}$ represents loop momenta perpendicular
to $z$-direction.
When $\delta z=0$, a local operator case, contributions from the
sail- and tadpole-type diagrams vanish, and the vertex-type reproduces
logarithmic UV and IR divergences of the local operator.
In the non-local case, the vertex-type is UV finite,
because the loop integral is regulated by $\delta z\not=0$,
while sail- and tadpole-type diagrams has logarithmic UV divergences.
Also, tadpole-type produces a linear UV divergence.
As we mentioned in the previous section this linear power divergence
should be subtracted.
The subtraction is carried out using the static potential, whose
one-loop expression is written as
\begin{eqnarray}
V(R)=-g^2C_F\frac{1}{4\pi R}+g^2C_F\int_{k_{\perp 0}}\frac{1}{k_{\perp 0}^2}
+{\cal O}(g^4),
\end{eqnarray}
where $k_{\perp 0}^2=k_1^2+k_2^2+k_3^2$.
From equations (\ref{EQ:subtracted_non-local_operator}) and
(\ref{EQ:subtraction_condition}),
it is clear that the linear divergence in the tadpole-type diagram and
the potential are canceled.
We can define a subtracted tadpole-type contribution as:
\begin{eqnarray}
I_{\rm tadpole}^{\rm subt}(\delta z)&=&
\frac{(4\pi)^2}{2}\int_{k_{\perp z}}
\left[\frac{1}{k_{\perp z}^3}
-\frac{1}{k_{\perp z}^3}e^{-k_{\perp z}|\delta z|}\right]. 
\end{eqnarray}

At this stage, we introduce a UV cutoff as a regulator.
Although the loop integrals are now three-dimensional,
the two-dimensional UV cutoff is enough to regulate the UV divergences.
The two directions for the cutoff correspond to usual transverse
direction in the Minkowski space.
Let $\mu$ be the two-dimensional UV cutoff scale and $\lambda$ be
the IR regulator, the loop integrals yield:
\begin{eqnarray}
I_{\rm vertex}(\delta z=0)=2\ln\frac{\mu}{\lambda},\;\;\;
I_{\rm sail}(\delta z=0)=0,\;\;\;
I_{\rm tadpole}^{\rm subt}(\delta z=0)=0,
\label{EQ:I_cont_deltaz_zero}
\end{eqnarray}
and for $\delta z\not=0$:
\begin{eqnarray}
I_{\rm vertex}(\delta z\not=0)&=&
-\int_{-\infty}^{\infty}dk_0
\left.\left(k_{\perp}+\frac{1}{\sqrt{k_0^2+1}}\right)
e^{-\sqrt{k_0^2+1}k_{\perp}}
\right|_{k_{\perp}=\lambda|\delta z|}^{\mu|\delta z|},
\label{EQ:I_cont_deltaz_nonzero_1}
\\
I_{\rm sail}(\delta z\not=0)&=&
4\ln\frac{\mu}{\lambda}
+2\int_{-\infty}^{\infty}dk_0
\left.
\frac{e^{-\sqrt{k_0^2+1}k_{\perp}}}{\sqrt{k_0^2+1}}
\right|_{k_{\perp}=\lambda|\delta z|}^{\mu|\delta z|},
\label{EQ:I_cont_deltaz_nonzero_2}
\\
I_{\rm tadpole}^{\rm subt}(\delta z\not=0)&=&
4\ln\frac{\mu}{\lambda}
+2\int_{-\infty}^{\infty}dk_0
\left.\left(\frac{e^{-\sqrt{k_0^2+1}k_{\perp}}}{\sqrt{k_0^2+1}}
+k_{\perp}{\rm Ei}\left[-\sqrt{k_0^2+1}k_{\perp}\right]\right)
\right|_{k_{\perp}=\lambda|\delta z|}^{\mu|\delta z|}.\nonumber\\
\label{EQ:I_cont_deltaz_nonzero_3}
\end{eqnarray}

\section{One-loop perturbative matching between continuum and lattice}

\begin{figure}[t]
\vspace*{-2mm}  
\begin{center}
\parbox{48mm}{
\begin{center}
\includegraphics[scale=0.58, viewport = 0 0 230 480, clip]
{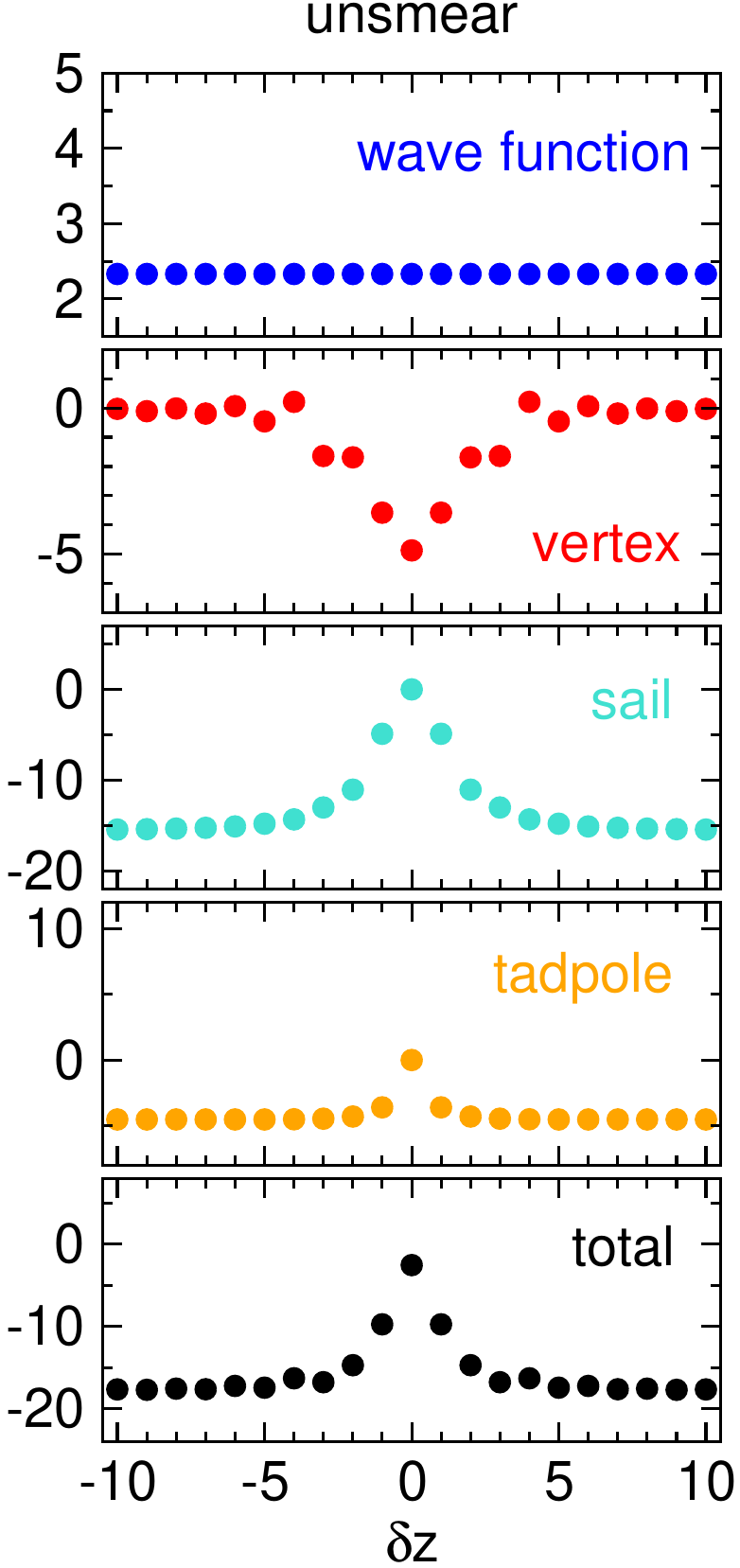}
\end{center}
}
\parbox{48mm}{
\begin{center}
\includegraphics[scale=0.58, viewport = 0 0 230 480, clip]
{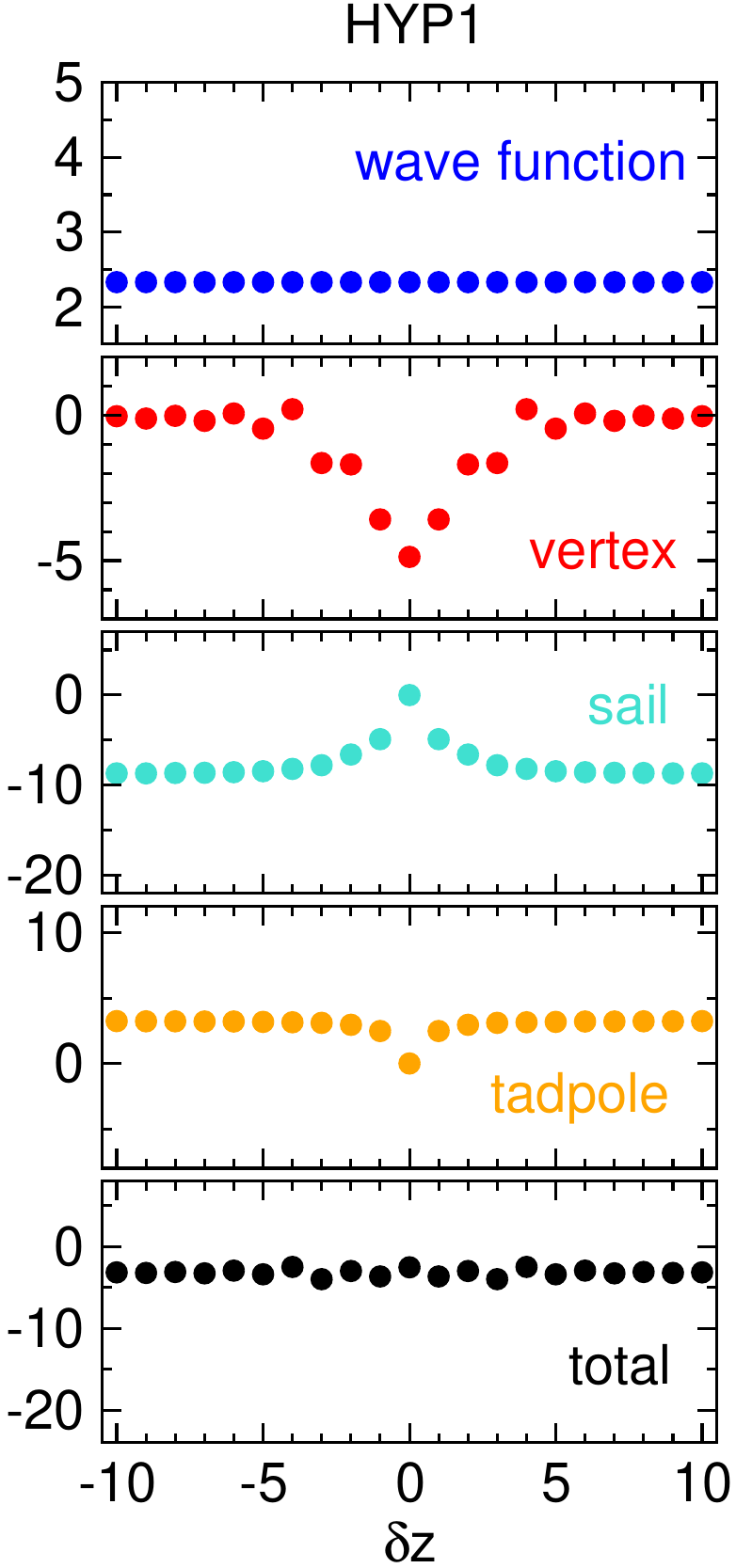}
\end{center}
}
\parbox{48mm}{
\begin{center} 
\includegraphics[scale=0.58, viewport = 0 0 230 480, clip]
{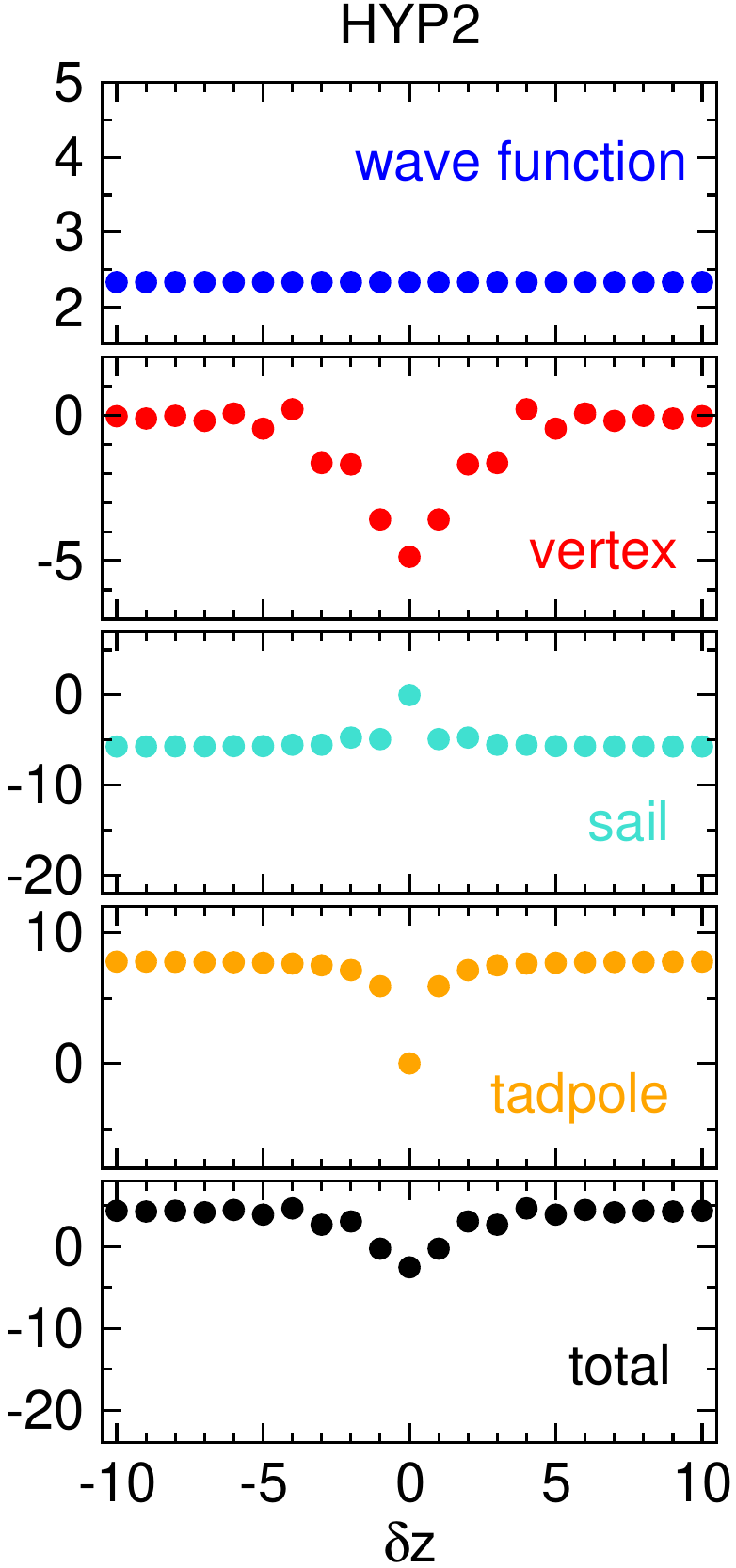} 
\end{center}
}
\vspace*{-2mm} 
 \caption{One-loop matching coefficients for each diagram:
 quark self-energy, vertex-type, sail-type and tadpole-type,
 as well as their total contribution.
 The linear divergence is subtracted, and the MF improvement is used.
 Three cases of gluon link smearing are considered for a Wilson line in the
 non-local operator: unsmear (left), HYP1 (center) and  HYP2 (right).}
 \label{FIG:one-loop_matching_ coefficients} 
\end{center}
\vspace*{-4mm}
 \end{figure}
In this section, we calculate the matching factor of the power divergence
subtracted non-local operator (\ref{EQ:subtracted_non-local_operator})
between continuum and lattice at the one-loop level.
The matching is done at each distance scales $\delta z$,
hence the matching factor could depend on $\delta z$.
With the multiplicative renormalization in equation
(\ref{EQ:non-local-bilinear_renormalization}),
we have the following matching pattern:
\begin{eqnarray}
O_{\rm cont}^{\rm subt}(\delta z)=Z(\delta z)O_{\rm latt}^{\rm subt}(\delta z).
\end{eqnarray}
In the following, we take a two-dimensional UV cutoff in the continuum
as mentioned in the previous section,
and the cutoff scale is set to be $\mu=a^{-1}$ (lattice cutoff).
For the lattice side, the na\"{i}ve fermion for the lattice perturbative
calculation is employed just for making the calculation simple.
Extending this work to other practical lattice fermions, such as
Wilson and domain-wall fermions, is straightforward, but just introduces
complications.
We also introduce link smearings for the Wilson line operator 
in the definition of the non-local operator for the lattice side.
The link smearing is often used for improving the S/N in the simulation,
and is also known to reduce power divergences.
We adopt two types of smearing, HYP1 and HYP2 in this study.
To improve convergence in the coupling expansion in the lattice
perturbative calculation, the mean-field improvement (MF) program is employed
(See reference \cite{Ishikawa:2016znu} for the details).
For this matching, the final result does not depend on 
the choice of the power divergence subtraction condition
(\ref{EQ:subtraction_condition}), because the relevant term
to the choice is canceled between continuum and lattice.
At one-loop level, the matching coefficient can be obtained by
taking the differences of the loop integrals between continuum
and lattice calculations:
\begin{eqnarray}
\delta I(\delta z)=I_{\rm cont}(\delta z)-I_{\rm latt}(\delta z),
\end{eqnarray}
where $I$ stands for integrals (\ref{EQ:I_cont_deltaz_zero}),
(\ref{EQ:I_cont_deltaz_nonzero_1}), (\ref{EQ:I_cont_deltaz_nonzero_2})
and (\ref{EQ:I_cont_deltaz_nonzero_3}) for continuum, and their
lattice counterparts.
The wave function renormalization is also included in the matching.

The one-loop matching coefficients are shown in
figure \ref{FIG:one-loop_matching_ coefficients},
separating contributions from each diagram.
The linear divergence is subtracted, then the $\delta z$ dependence in
the large $\delta z$ region is flat.
This result is consistent with an intuition that 
the difference of the continuum and the lattice is of only the UV structure.
The Wilson link smearing gives a tiny one-loop total coefficient
compared to the unsmeared case.
This small coefficient is preferable for perturbative accuracy.

\section{Summary}

We reported our effort to address two of the major uncertainties in 
extracting PDFs from quasi-PDFs calculated on the lattice:
the power divergences and matching between continuum and lattice.
Since the power divergences must be subtracted nonperturbatively,
we presented the subtraction scheme using a static quark potential.
We also derived the one-loop matching factor 
between the continuum and lattice calculations.
Although other nonperturbative renormalization techniques, 
such as RI/MOM scheme, might be preferable for better accuracy in 
computing the matching, our one-loop perturbative calculation could provide
a good guidance for the future efforts.

\section*{Acknowledgements}

This work is supported in part by the U.S. Department of Energy,
under contract DE-AC05-06OR23177.


\end{document}